# Topological valley currents via ballistic edge modes in graphene superlattices near the primary Dirac point

Yang Li[1,2*], Mario Amado[1,3], Timo Hyart[4,5], Grzegorz P. Mazur[4], and Jason W. A. Robinson[1*]

[1]Department of Materials Science & Metallurgy, University of Cambridge, 27 Charles Babbage Road, Cambridge CB3 0FS, United Kingdom.

[2]Cambridge Graphene Centre, University of Cambridge, 9 JJ Thomson Avenue, Cambridge CB3 0FA, United Kingdom.

[3]Nanotechnolgy Group, Department of Physics, University of Salamanca, 37008 Salamanca, Spain.

[4]International Research Centre MagTop, Institute of Physics, Polish Academy of Sciences, Aleja Lotników 32/46, PL-02668 Warsaw, Poland.

[5]Department of Applied Physics, Aalto University, 00076 Aalto, Espoo, Finland.

**Abstract**

Graphene on hexagonal boron nitride (hBN) can exhibit a topological phase via mutual crystallographic alignment. Recent measurements of nonlocal resistance ($R_{nl}$) near the secondary Dirac point (SDP) in ballistic graphene/hBN superlattices have been interpreted as arising due to the quantum valley Hall state. We report hBN/graphene/hBN superlattices in which $R_{nl}$ at SDP is negligible, but below 60 K approaches the value of $h/2e^2$ in zero magnetic field at the primary Dirac point with a characteristic decay length of 2 µm. Furthermore, nonlocal transport transmission probabilities based on the Landauer-Büttiker formalism show evidence for spin-degenerate ballistic valley-helical edge modes, which are key for the development of valleytronics.

*E-mail: yl539@cam.ac.uk; jjr33@cam.ac.uk

**Introduction**

Graphene offers opportunities for fundamental solid-state physics and applications in spintronics, because it can support currents with charge, spin, and valley degrees of freedom[1]. Graphene valley-dependent electronics was proposed a decade ago[2,3], but inversion symmetry in graphene makes it challenging to apply the valley degree of freedom in electronics. However, minimal lattice mismatch[4] (1.8%) between graphene and hexagonal boron nitride (hBN) results in a rotation-dependent moiré pattern, which leads to weak periodic potentials[5] and broken inversion symmetry[6] in graphene. In atomically aligned graphene/hBN, a band gap at the primary Dirac point (DP, $V_D$) is formed[6-8] and secondary Dirac points (SDPs) are stabilized at energy relating to the moiré wavelength[4,9].

Recently, nonlocal resistances ($R_{nl}$) in aligned graphene/hBN Hall bars at the DP (or SDP) and in zero magnetic field have been interpreted as being related to a finite Berry curvature, which leads to the valley Hall effect (VHE) due to a coupling between the valley and the electron orbital motion[3,8,10,11]. In Ref. [8], $R_{nl}$ is around 1 kΩ at the DP in both encapsulated (i.e. hBN/graphene/hBN) and non-encapsulated (i.e. hBN/graphene/SiO$_2$) Hall bars. In Ref. [10], $R_{nl}$ reaches the quantum-limited value at the SDP in a ballistic encapsulated Hall bar, albeit with an anomalously low $R_{nl}$ at the DP (similar value to Ref. [8]). In the Hall bars reported in Ref. [8], the top hBN layer is misaligned with respect to the graphene by 10°, whilst in Ref. [10] the top and bottom hBN layers are aligned to graphene. The alignment seems to have a large effect on both the band gap and ballistic character of the nonlocal transport. Here we report $R_{nl}$ measurements in encapsulated graphene Hall bars (hBN/graphene/hBN) with different alignment angles and focus on nonlocal transport near the DP where $R_{nl}$ approaches $h/2e^2$ in zero magnetic field below 60 K, in contrast to Refs. [8] and [10].

**Results**

**Fabrication and characterization of graphene superlattices.** hBN/graphene/hBN Hall bars are fabricated by van der Waals assembly with side-contacts (see Methods section for further details) as shown in Fig. 1a, b. The relative rotation angle ($\varphi$) between hBN and graphene is determined using a transfer system with a rotating stage under an optical microscope with an accuracy of better than 1.5⁰. Local and nonlocal transport measurements are taken across a range of temperatures (8.8-300 K) and magnetic fields (0-2.5 T). We focus on the electronic transport of three types of devices denoted as I, II, and III (for device mobility characterization see Supplementary Note 1). For device I (a field-effect mobility of $\mu \approx$ 220,000 cm$^2$ V$^{-1}$ s$^{-1}$ at 9 K, Fig. 1a and Supplementary Figure 2a), the graphene is aligned to both top and bottom hBN layers ($\varphi \approx 0$⁰). For device II ($\mu \approx$ 350,000 cm$^2$ V$^{-1}$ s$^{-1}$ at 9 K, Supplementary Figure 1a and 2b), the graphene is aligned to the top layer of hBN ($\varphi \approx 0$⁰) but misaligned ($\varphi \approx 30$⁰) with respect to the bottom hBN. For device III ($\mu \approx$ 50,000 cm$^2$ V$^{-1}$ s$^{-1}$ at 9 K, Supplementary Figure 1b and 2c), the graphene is misaligned with respect to the top ($\varphi \approx 10$⁰) and bottom ($\varphi \approx -10$⁰) hBN layers.

For all Hall bars investigated, Raman spectroscopy is performed over the entire area of graphene to confirm structural uniformity post transfer (Fig. 1c, Supplementary Note 1 and Supplementary Figure 1c-e). We do not observe a D-peak near 1345 cm$^{-1}$, ruling out detectable lattice defects in graphene. Hall bars are fabricated in areas where the full width at half maximum of the 2D-peak [FWHM(2D)] is in the 25-30 cm$^{-1}$ range (Fig. 1c). In Fig. 1d we show the 2D-peaks of the different structures investigated: FWHM(2D) of device I (27 cm$^{-1}$) is larger than devices II and III (17 cm$^{-1}$ and 22 cm$^{-1}$) for which $\varphi$ are 30⁰ and 10⁰. The Raman 2D-peak of graphene is sensitive to $\varphi$, and FWHM(2D) increases by rotating from a misaligned position to an aligned position, which is due to a strain distribution with matching moiré potential periodicity[12].

**Local electrical transport properties.** Figure 2a shows local measurements in zero magnetic field at 9 K with a pronounced peak in $\rho_{xx}$ at the primary DP. Two additional peaks are symmetrically visible on both sides of the DP in the higher carrier density regime. The appearance of SDP depends on moiré minibands occurring near the edges of the superlattices Brillouin zone[4,9] and the moiré wavelength $\lambda$ of device I is calculated to be around 10 nm ($\varphi < 1°$) (see Supplementary Note 2) and for devices II and III, there are no SDPs within the gate voltage range investigated (±20 V) as the $\varphi > 10°$ (requires $|V_{TG}-V_D| > 100$ V). Device I also shows the ballistic character (see Supplementary Note 3).

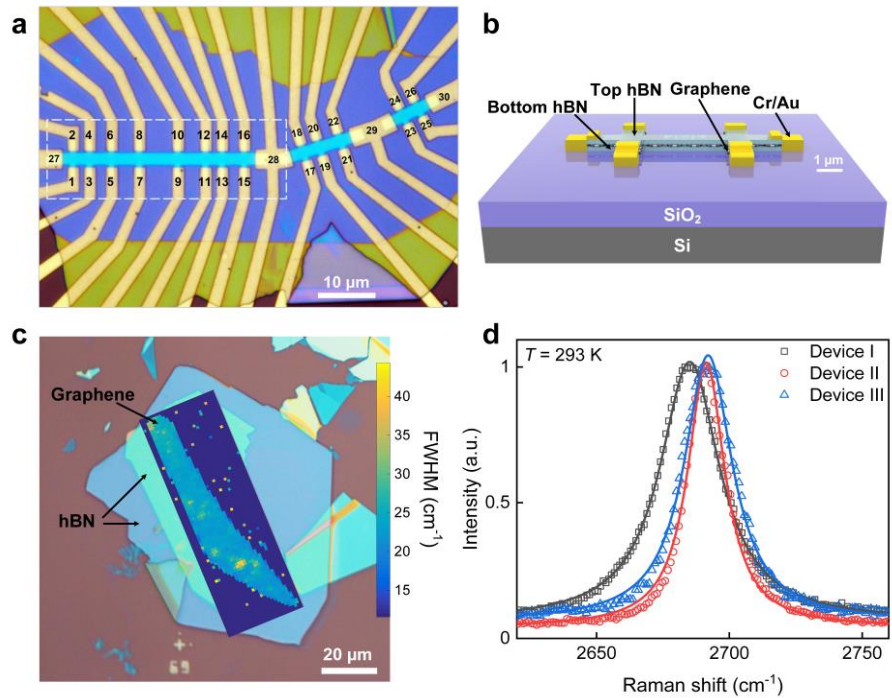

**Fig. 1 hBN/graphene/hBN Hall bar structure and Raman properties. a** Optical micrograph of device I in the white square. **b** Schematic illustration of a device. **c** Raman spectra map (dark blue rectangle region) at 293 K for device I showing a high degree of uniformity in the full width at half maximum (FWHM) of the graphene 2D-peak. **d** Raman 2D-peak positions of graphene with Lorentzian fits for devices I, II and III.

In Fig. 2b, c, when temperature is decreasing, a $v = 0$ plateau appears in $\sigma_{xy}$ and a double-peak structure appears in $\sigma_{xx}$ when the gate voltage ($V_{TG}$) is close to the DP (see Supplementary Note 4 for detailed discussions). Two different types of conductivity variations are seen in Fig. 2b: one is insulating meaning $\sigma_{xx}$ decreases at lower temperatures when $V_{TG}$ is close to the DP; the other is metallic in which $\sigma_{xx}$ increases at lower temperatures. The critical point separating these two regimes is the crossing point of all the curves measured at different temperatures, where $\sigma_{xx}$ is independent of temperature ($T$) indicating quantum Hall state transitions. Figure 2d, e show the evolutions of $\rho_{xx}$ and $\sigma_{xy}$ with $V_{TG}$ and increasing perpendicular magnetic fields ($B$). Standard quantum Hall state with plateaux in $\sigma_{xy}$ and zeros in $\sigma_{xx}$ at filling factors $v = \pm 2, \pm 6, \pm 10$ ... is observed. A striking feature is the insulating region near the DP with increasing $B$, where $\rho_{xx} \geq h/2e^2$.

A quantum Hall effect gap at the DP in hBN/graphene/hBN superlattices occurs due to electron interactions and broken sublattice symmetry[6,13]. From the insulating behavior of $\sigma_{xx}$ at the DP, we fit an Arrhenius function $\sigma_{xx}(T^{-1})$ to estimate the band gap in Fig. 2f. The thermally excited transport exhibits two distinct regimes of behavior, separated by a characteristic temperature, which we define as $T^*$. For $T > T^*$, transport is thermally activated[10] meaning that $\sigma_{xx,min} \propto exp(-E_a/2k_BT)$, where $k_B$ is the Boltzmann constant and $E_a$ is the band gap energy. The band gap is estimated to be 391.2 ± 21.8 K (33.7 ± 1.9 meV) for device I and 210.1 ± 11.2 K (18.1 ± 1.0 meV) for device II. The larger band gap at the DP for device I can be associated with the commensurate state, because the whole area of graphene, which is aligned to hBN, would have the same crystal structure as hBN and tend to increase the gap. Device II has a smaller $E_a$ due to a suppression of the commensurate state by one of the misaligned hBN layers[7]. For $T < 60$ K, $\sigma_{xx,min}$ decreases slowly with lower $T$ than the activated transport indicating that in this regime the effect of the thermally activated bulk carriers can be neglected. In Fig. 2b, the temperature dependence of the longitudinal conductivity in 2 T shows the appearance of $v = 0$ state below $T < 60$ K. The appearance of the quantum Hall state also requires that the effect of the thermally activated bulk carriers can be neglected.

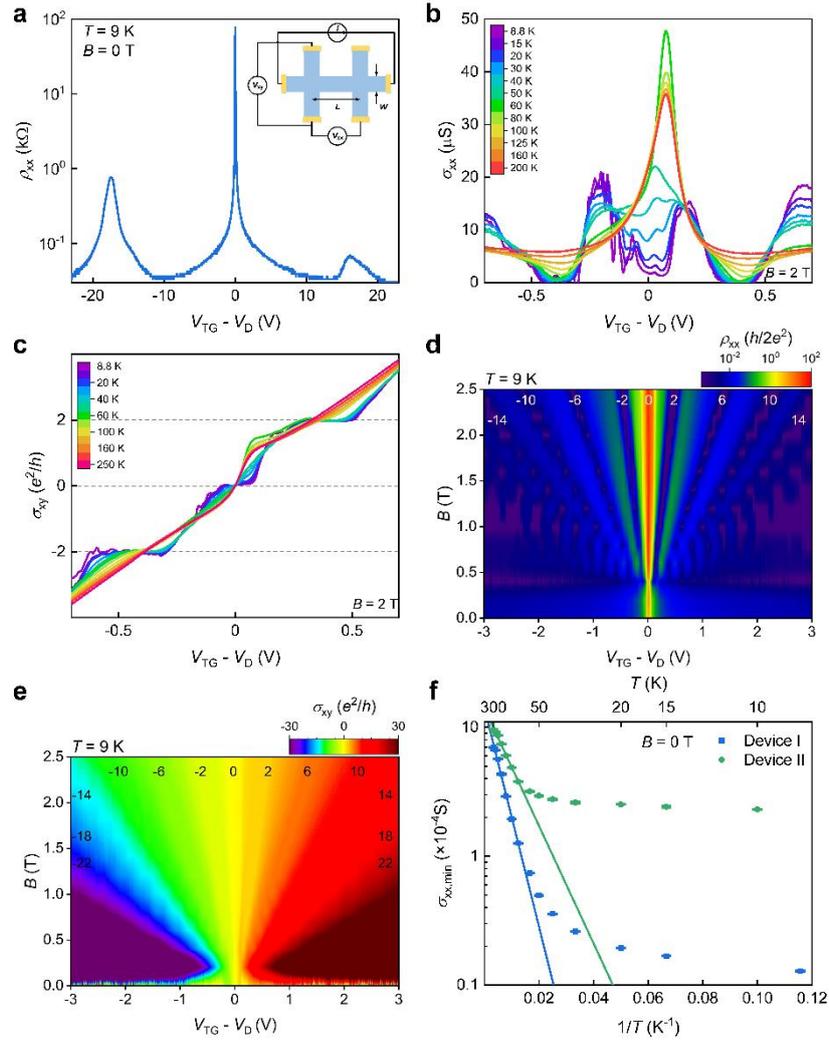

**Fig. 2 Local transport in hBN/graphene/hBN Hall bar. a** longitudinal resistivity ($\rho_{xx}$) vs gate voltage ($V_{TG} - V_D$) in zero magnetic field ($B = 0$) at 9 K, which shows pronounced peaks at the primary Dirac point (DP, $V_{TG} = V_D$) and the secondary Dirac points ($V_{TG} - V_D = \pm16.75$ V). Inset shows schematic illustrations of the local measurement setup, where $L$ is the distance between the current path and voltage probes, and $W$ is the device width. **b**, **c** Temperature dependence of longitudinal conductivity ($\sigma_{xx}$) and Hall conductivity ($\sigma_{xy}$) vs $V_{TG} - V_D$ in 2 T. Color scale shows the temperatures from 8.8 K to 250 K. **d**, **e** Quantization in $\rho_{xx}$ and $\sigma_{xy}$ vs $V_{TG} - V_D$ and $B$ at 9 K. $B$ applied perpendicular to the device. Numbers denote filling factors for the quantum Hall states extending from the DP. Color scale in **d** shows the $\rho_{xx}$ in the unit of $h/2e^2$, and in **e** shows the $\sigma_{xy}$ in the unit of $e^2/h$. **f** Arrhenius plot of $\sigma_{xx}$ at the DP for devices I and II, where the error bars come from the gate voltage sweeping step. All measured results are shown from device I.

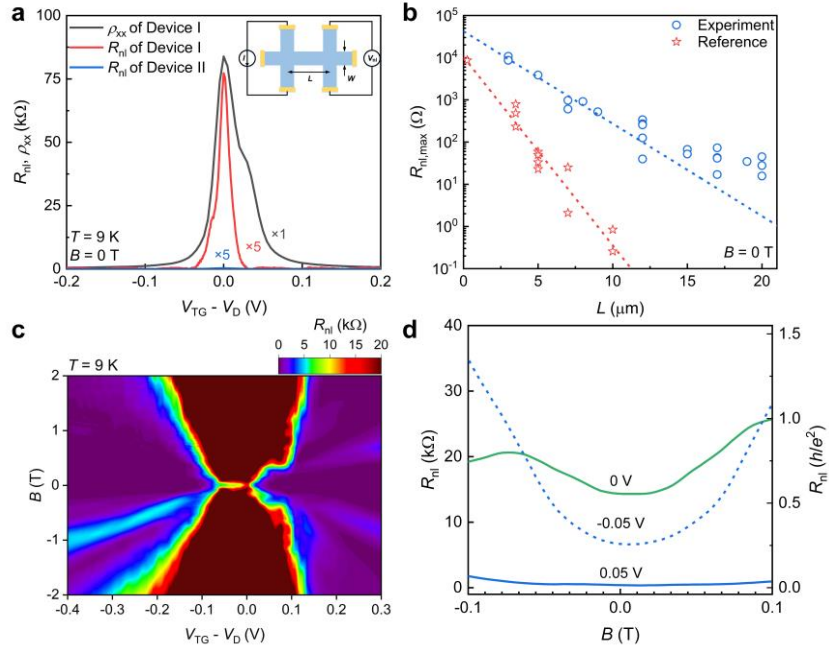

**Fig. 3 Long-range nonlocal transport in hBN/graphene/hBN Hall bar. a** longitudinal resistivity ($\rho_{xx}$) and nonlocal resistance ($R_{nl}$) vs gate voltage ($V_{TG} - V_D$) in devices I and II (labelled) with $L = 3$ µm. Inset shows schematic picture of the nonlocal measurement setup, where $L$ is the distance between the current path and voltage probes, and $W$ is the device width. **b** Distance dependence for $R_{nl}$ in device I. $R_{nl}$ decays exponentially with increasing distance at 9 K. $L$ varies from 3 µm to 20 µm and $W$ is 2 µm. Reference data is taken from Ref. [8]. **c** $R_{nl}$ vs $V_{TG} - V_D$ and magnetic field ($B$) at 9 K. Color bar shows the $R_{nl}$ from 0 to 20 kΩ. **d** $R_{nl}$ vs $B$ for different $V_{TG} - V_D$.

**Nonlocal electrical transport properties.** In Fig. 3a, $R_{nl}$ for device I shows a sharp peak (15.45 kΩ) at the DP and $\rho_{xx}$ has a $1/n$ dependence, which decreases at a slower rate than $R_{nl}$ over the entire range of $V_{TG}$ investigated. Within achievable $V_{TG}$ (±30 V), we do not observe nonlocal transport at the SDPs which could be due to charge inhomogeneity suppression[8]. $R_{nl}$ for device II (same geometry as device I) is smaller (60 Ω), consistent with a misalignment between graphene and hBN. In addition, the improved electronic properties of graphene on hBN enable long-range topological valley currents[8,10]. $R_{nl}$ exponentially decays as a function of nonlocal distance ($L$) in graphene [i.e. $R_{nl} \propto exp(-L/\xi)$] with a characteristic length $\xi \approx 2.0$ µm (Fig. 3b). The maxima in $R_{nl}$ for all values of $L$ investigated are at least an order of magnitude larger than previously reported values in equivalent devices with similar mobility[8]. With $B$ applied, we observe a rapid broadening and increase in the $R_{nl}$ peak above 0.1 T (Fig. 3c, d) due to contribution from charge-neutral spin currents, which become appreciable with broken time reversal symmetry[8,14]. In addition, $R_{nl}$ under magnetic field can have contributions from heat current and the quantum Hall effect edge current[8,10].

**Discussion**

There are several possible explanations for the existence of a finite $R_{nl}$ in zero magnetic field, including ohmic and thermal effects, charge-neutral spin current, VHE, edge-modes-driven quantum valley Hall state and one-dimensional conducting channel. Ohmic contributions account for around 1% of $R_{nl}$ as demonstrated in Supplementary Figure 8. The carrier density dependence, which follows $\rho_{xx}$, is incompatible with the observed nonlocal response (Fig. 4a). Thermal contributions to $R_{nl}$ in device I are ruled out (see Supplementary Note 5). Devices I and II show different magnitudes of $R_{nl}$ at the DP, which cannot be explained by charge-neutral spin current, as these are indifferent to the relative alignment angle and require broken time reversal symmetry[14,15]. VHE is induced by the accumulated Berry curvature near hot spots and is associated with the transverse valley Hall conductivity ($\sigma_{VH}$), which can be detected via $R_{nl}$ according to $R_{nl} \propto (\sigma_{VH})^2 \rho_{xx}^3$ (in the limit of $\sigma_{VH} \ll 1/\rho_{xx}$) from the semiclassical transport theory[8,16-18]. $R_{nl} \propto \rho_{xx}^3$ holds for $T \geq 60$ K (Fig. 4b and Supplementary Note 6), but not for $T < 60$ K at both the hole and electron sides (Fig. 4c) where $R_{nl}$ starts to show deviations from the semiclassical transport theory and approaches $h/2e^2$ in the insulating regime at the DP. If the quantum-limited value of $R_{nl}$ is due to the edge states, we also have the requirement that the effect of the thermally activated bulk carriers can be neglected ($T < 60$ K). However, in the case of large valley Hall conductivity[19], $R_{nl}$ from the bulk valley Hall effect may become independent of $\rho_{xx}$. The bulk valley Hall effect can lead to a current distribution localized in the vicinity of the edge and in many respects resembles the edge transport[20]. For our devices, $R_{nl}$ is not showing a smooth and monotonic dependence and saturation as a function of $\rho_{xx}$ expected in the case of a bulk response, but rather $R_{nl}$ exhibits fluctuations, which resemble the mesoscopic conductance fluctuations in the quasiballistic regime of a quantum wire as expected in the case of unprotected edge modes.

To confirm the origins of $R_{nl}$ for $T < 60$ K, we first measure $R_{nl}$ using a six-terminal configuration (Fig. 4d, e). In Ref. [10], a quantum valley Hall state from a pair of valley-helical counter-propagating edge modes is proposed in order to explain the quantum-limited values of $R_{nl}$ – i.e. by assuming the minimal model where a pair of ballistic counter-propagating edge modes connect terminals, the theoretical values for the quantized $R_{nl}$ can be calculated using Landauer-Büttiker formalism[21,22]. In device I, theoretical values of $R_{nl}$ based on the minimal model for (*I*: 11,12, *V*: 13,14) and (*I*: 11,12, *V*: 14,28) are $2h/3e^2$ and $h/3e^2$, respectively, which are compared with the experimental results shown in Fig. 4d. While valley-helical edge modes exist for specific type of edges[11,23,24], they are spin-degenerate in all proposed theoretical models due to spin-rotation symmetry, and therefore the microscopic origin of the quantum valley Hall state in Ref. [10] remains unknown.

To investigate the origin of quantum valley Hall state, we measure $R_{nl}$ systematically using a ten-terminal configuration (Fig. 4e) in order to determine the transmission matrix. Device I is fabricated with eighteen terminals (Fig. 1a), fourteen of which show relatively low contact resistances (Fig. 4e). We select ten terminals located symmetrically to measure. The calculated transmission matrix based on the Landauer-Büttiker formalism (see Supplementary Note 7) does not agree with the minimal model for edge mode transport proposed in Ref. [10]. However, when the ballistic counter-propagating edge modes enter these unused but connected terminals, they interact with a reservoir of states and equilibrate to the chemical potential determined by the voltage at each terminal. Therefore, electrons will be injected backward and forward with equal probability. These unused terminals in-between the measured terminals effectively reduce the ideal transmission probability by a half. The transmission probabilities are approximately 2.0 between terminals 12 and 14, and reach 1.0 between terminals 5 and 27, 6 and 27, 14 and 28 in the narrow $V_{TG}$ range near the DP, consistent with spin-degenerate ballistic counter-propagating edge mode transport. In addition, it is known that commensurate stacking in aligned van der Waals heterostructure ($\varphi < 1°$) leads to the soliton-like narrow domain walls[25]. One-dimensional conducting channels exist at these domain walls, which can form a network leading to $R_{nl}$ when bulk graphene/hBN superlattices domains become insulating[26–30]. If edge modes intersect with domain walls, the electrons can go into two different directions at the intersection, and this will lower the transmission probabilities for each given direction. The transmission probabilities for terminals 5, 6, 7 and 8 (Supplementary Table 1) are significantly smaller than expected values based on the ballistic edge mode transport, which perhaps are consistent with the existence of domain walls.

To summarize, we have investigated hBN/graphene/hBN Hall bars with a field-effect mobility of 220,000 cm$^2$ V$^{-1}$ s$^{-1}$ at 9 K and low charge impurities. Alignment between hBN and graphene ($\varphi < 1°$) leads to a 33.7 meV band gap at the DP. In zero magnetic field and 9 K, a $v = 0$ state in gapped graphene is demonstrated in $\sigma_{xx}$ and $\sigma_{xy}$ with large $R_{nl}$ values close to $h/2e^2$. $R_{nl}$ decays nonlocally over distances of 15 μm with a characteristic constant of 2 μm. Nonlocal measurements suggest that, below 60 K a spin-degenerate ballistic counter-propagating edge mode is dominant, and there is a possible secondary contribution from a network of one-dimensional conducting channels appearing at the soliton-like domain walls. A further direct imaging of the edge modes[31-33] would be desirable for conclusive determination of the mechanism of nonlocal transport. The valley-helical ballistic edge modes offer important possibilities for electronic applications beyond quantum spin Hall effect and quantum anomalous Hall effect since quantized resistance can be observed at higher temperature with a tunable energy gap through valley coupling.

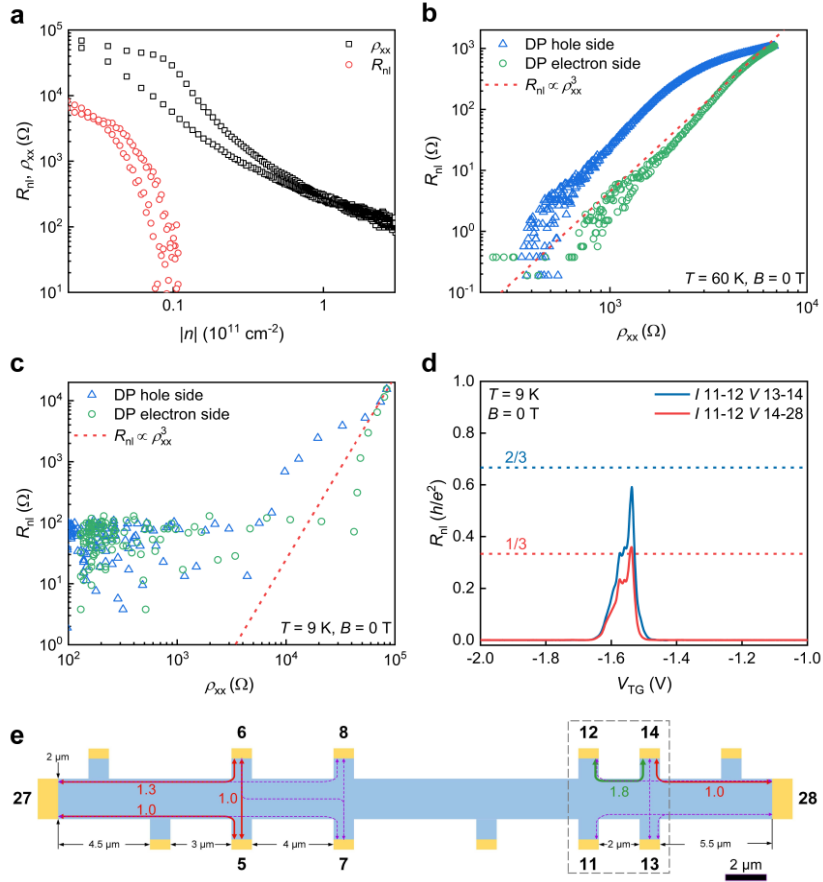

**Fig. 4 The origins of nonlocal resistance ($R_{nl}$). a** Logarithmic scale of longitudinal resistivity ($\rho_{xx}$) and $R_{nl}$ vs carrier density ($n$). **b, c** Scaling of $R_{nl}$ vs $\rho_{xx}$ at temperature ($T$) of 60 K and 9 K, without applied magnetic field ($B$), where the DP is the Dirac point. **d** $R_{nl}$ in a six-terminal configuration with channel length of 3 μm at 9 K vs gate voltage ($V_{TG}$), where the horizontal (dotted) lines show theoretical values based on the minimal model. **e** Nonlocal transport geometry for the six-terminal (black square) and ten-terminal measurements. Six-terminal configuration includes terminals 11, 12, 13, 14, 27 and 28. Ten-terminal configuration includes terminals 5, 6, 7, 8, 11, 12, 13, 14, 27 and 28. Transmission probabilities ($Tr_{ij}$) are indicated with solid arrows for $Tr_{ij} \geq 1.0$ and with dashed arrows for $0.5 \leq Tr_{ij} < 1.0$. Several pairs of terminals show the expected values for the ballistic edge mode transport. The error bars for all measured resistances come from the gain accuracy of lock-in amplifiers, which are ±0.4%.

## Methods

**Device fabrication.** Mechanically exfoliated graphene is transferred to high quality single crystals of hBN (purchased from HQ Graphene) which are exfoliated on top of an oxidized Si wafer (285 nm $SiO_2$) and then covered with another hBN crystal. To align the crystal lattices, straight and long edges of graphene and hBN flakes indicate the principal crystallographic direction, which are selected by optical microscope. During the assembly, the bottom hBN crystal is rotated relatively to the graphene to make their edges parallel. The top hBN crystal is again transferred by using the same method and then carefully aligned to the bottom graphene/hBN stack. Electron beam lithography and reactive ion etching are then employed to define multi-terminal Hall bars.

**Measurement setup.** Transport measurements are performed using lock-in amplifiers at low frequency (7 Hz) with low excitation currents (1-10 nA at 9 K) as functions of magnetic field (0-2.5 T) and top-gate bias at different temperatures (8.8-300 K). A series resistance of 100 MΩ is introduced to maintain a constant current condition that is confirmed by the signal from the lock-in amplifier, which measures the current fed through a 10 kΩ resistor in series. For the local measurement (Supplementary Figure 3a), a current $I$ is applied between the contacts (5 and 6), the measured voltage between the contacts (1 and 2) is Hall voltage ($V_{1,2}$) and between the contacts (2 and 4) is longitudinal voltage ($V_{2,4}$). Then $\rho_{xx}$ is given by $V_{2,4}/I$ divided by the geometrical factor, $L/W$. The Hall resistivity $\rho_{xy}$ is calculated by $\rho_{xy} = V_{1,2}/I$. For the nonlocal measurement (Supplementary Figure 3b), a current $I$ is applied between the contacts (1 and 2), the measured voltage between the contacts (3 and 4) is nonlocal voltage ($V_{3,4}$) and is often converted to resistance by dividing the injection current ($R_{nl} = V_{3,4}/I$).

**Data Availability**

The data that support the findings of this study are available from the corresponding author upon reasonable request.

**Acknowledgments**

The research was funded by the Royal Society and the EPSRC through an EPSRC-JSPS International Network Grant (EP/P026311/1). Y.L. was supported through China Scholarship Council (CSC) Cambridge International Scholarship and Cambridge Trust. M.A. was supported from the MSCA-IFEF-ST Marie Curie (Grant 656485-Spin3), the Agencia Estatal de Investigación of Spain (Grant MAT2016-75955), and the Junta de Castilla y León (Grant SA256P18). T.H. and G.P.M were supported by the Foundation for Polish Science through the IRA Programme co-financed by EU within SG OP. G.P.M. was supported by the National Science Center (Poland) through ETIUDA fellowship (Grant No. UMO-2017/24/T/ST3/00501).

**Author contributions**

J.W.A.R. conceived the experiment and directed the research. Y.L. fabricated the devices, performed the electrical, magnetic and structural measurements. T.H. developed the theory interpretation. M.A. and G.P.M. supported electrical measurements. Y.L., T.H., J.W.A.R. and M.A. wrote the paper. All authors commented on the manuscript.

**Competing interests**

The authors declare no competing interests.

**Additional information**

Supporting information is available for this paper, which includes analysis of extrinsic contributions to nonlocal signals, and transmission probability calculation.

**Supplementary Note 1. Device characterization**

To investigate structural and electronic homogeneity of the graphene on hBN, Raman spectroscopy measurements are performed over entire region at 293 K with a laser excitation at a wavelength of 532 nm, are shown in Fig. 1c and Supplementary Figure 1c,d. The fixed position spectra in Supplementary Figure 1e exhibit the characteristic hBN peak as well as the G- and 2D-peaks of graphene. The absence of a D-peak around 1345 cm$^{-1}$ indicates there is no significant lattice defects in the graphene. The positions of the G- and 2D-peaks are at 1580 – 1585 cm$^{-1}$ and 2686 – 2691 cm$^{-1}$, respectively, meaning an overall low doping concentration of graphene[1]. The quality of the device is characterized through field-effect mobilities: for device I it is in the 100,000-220,000 cm$^2$ V$^{-1}$ s$^{-1}$ range at 9 K in Supplementary Figure 2a, for device II it is in the 200,000-350,000 cm$^2$ V$^{-1}$ s$^{-1}$ at 9 K in Supplementary Figure 2b, and for device III it is in the 10,000-50,000 cm$^2$ V$^{-1}$ s$^{-1}$ at 9 K in Supplementary Figure 2c.

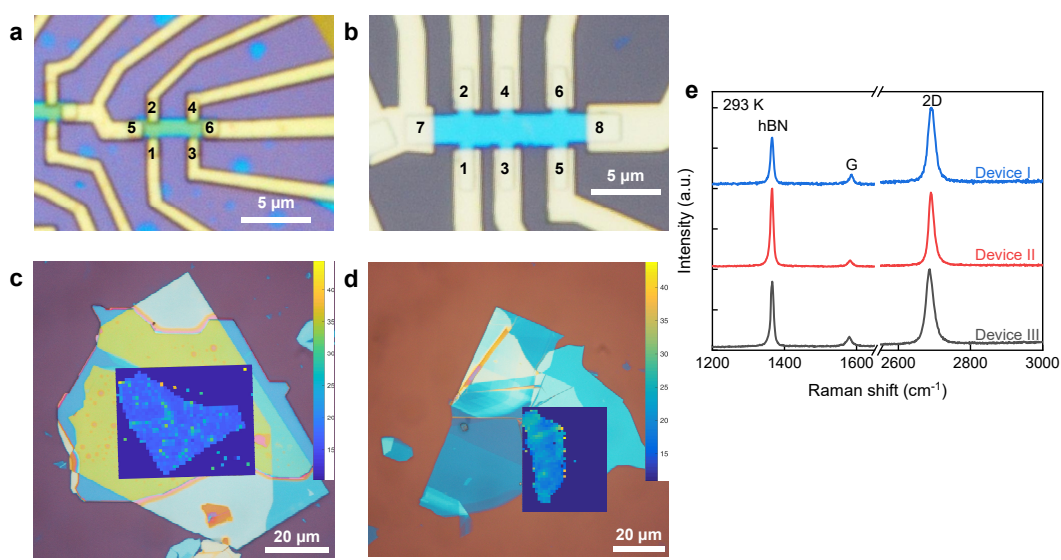

**Supplementary Fig 1. Structures and properties of devices II and III. a** Optical micrograph of device II. **b** Optical micrograph of device III. **c, d** The full width at half maximum of graphene 2D-peak [FWHM(2D)] Raman maps (dark blue rectangle region) of devices II and III, showing the encapsulated graphene. **e** Raman spectra at 293 K for all devices (labelled).

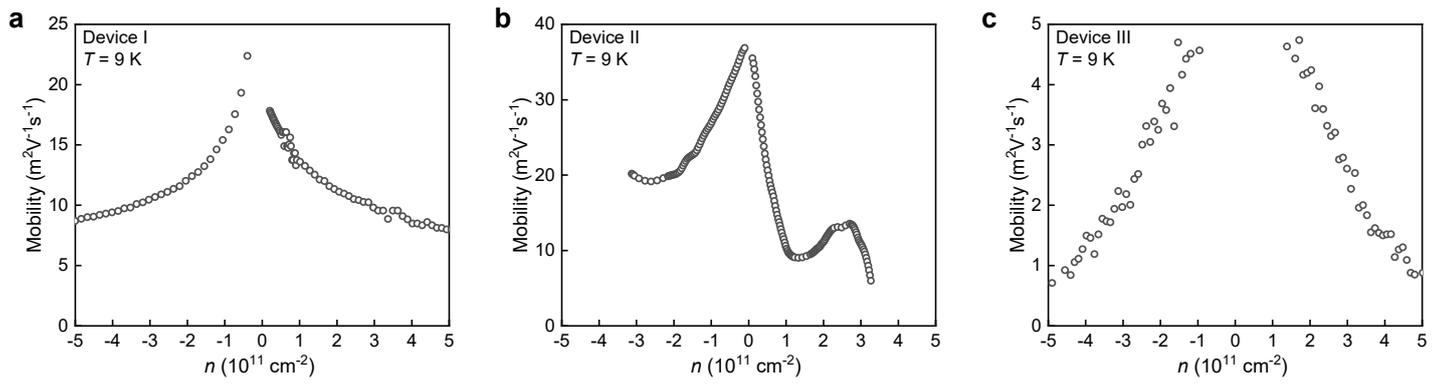

**Supplementary Fig. 2.** Field-effect mobility of hBN/graphene/hBN devices (labelled).

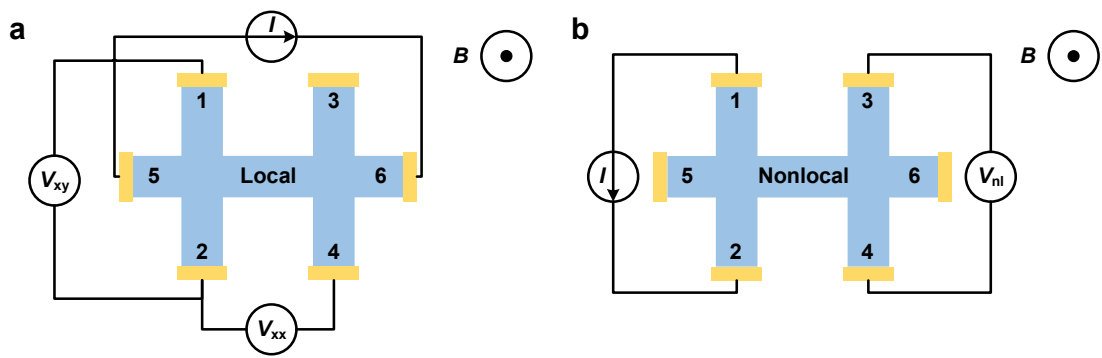

**Supplementary Fig. 3. Transport measurement configurations. a** Local measurement setup. **b** Nonlocal measurement setup.

**Supplementary Note 2. Moiré wavelength calculation**

hBN may have the same lattice structure as graphene with a 1.8% longer lattice constant. The alignment between the graphene and hBN lattices leads to moiré patterns. The moiré wavelength $\lambda$ is described as[2]

$$\lambda = \frac{(1+\delta)a}{\sqrt{2(1+\delta)(1-\cos\varphi)+\delta^2}} \tag{S1}$$

where $\delta$ is the lattice mismatch between hBN and graphene, $a$ is the graphene lattice constant, $\varphi$ is the relative rotation angle between hBN and graphene. Supplementary Figure 4 plots the $\lambda$ as a function of $\varphi$ and shows a maximum value of about 14 nm.

The appearance of SDP depends on moiré minibands, which occur near the edges of the superlattice Brillouin zone and are characterized by the energy[2,3] of SDP,

$$|E_{\text{SDP}}| = \sqrt{\pi|n|}v_F\hbar = \frac{2\pi\hbar v_F}{\sqrt{3}\lambda} \tag{S2}$$

where $n$ is the carrier density related to the SDP, $v_F$ is the Fermi velocity, $\hbar$ is Planck's constant divided by $2\pi$ and $\lambda$ is moiré wavelength. From Supplementary Equation (S2), the position of SDP corresponds to a carrier density of $n = 4\pi/3\lambda^2$, and in the case of $\varphi = 0°$, $\lambda = 14$ nm yields $n \approx 2 \times 10^{12}$ cm$^{-2}$.

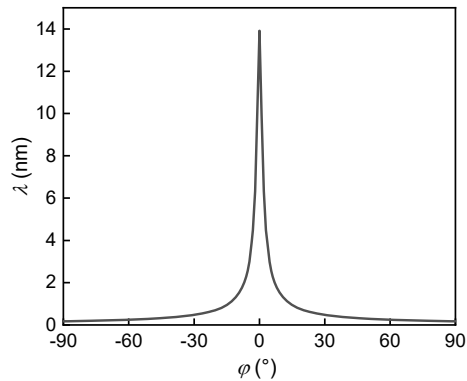

**Supplementary Fig. 4. Moiré wavelength as a function of the relative rotation angle ($\varphi$) between the graphene and hBN.**

## Supplementary Note 3. Mean free path

At all temperatures, diffusive transport prevails at very low carrier density near the DP, as the mean free path $L_{mfp}$ depends on $n$, which is determined by $L_{mfp} = \mu h \sqrt{n} / 2e\sqrt{\pi}$ (Supplementary Ref. [4]) ($h$ is Planck's constant, $n$ is carrier density, $e$ is electron charge and mobility $\mu$ can be calculated using Drude formula $\sigma = en\mu$). In Supplementary Figure 5a, $L_{mfp}$ increases with decreasing temperature and saturates to be 2 µm below 20 K, which is comparable to the device geometry. In device I, $R_{nl}$ has negative values for both electron and hole doping up to 60 K, with a distinct minimum in the hole regime close to the DP (Supplementary Figure 5b). The negative $R_{nl}$ behaves a strong temperature dependence and becomes positive with increasing temperature, indicating the transition to the regime of diffusive charge transport, dominated by electron-phonon scattering. A negative $R_{nl}$ indicates a ballistic overshoot of charge carriers from the injection contact into the detection contact[5]. The mean free path must exceed the minimum dimension of the device, which is 2 µm. All of these results confirm the ballistic character of device I.

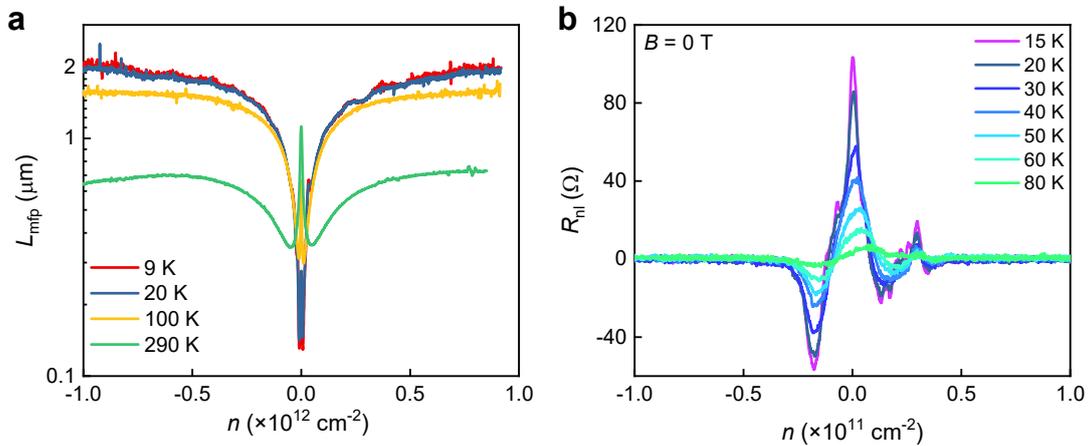

**Supplementary Fig. 5. Ballistic character of device I. a** Mean free path $L_{mfp}$ calculated from the diffusive regime as a function of the carrier density $n$ for different temperatures. **b** Negative nonlocal resistance $R_{nl}$ measured from 15 K to 80 K and zero magnetic field ($B = 0$) with channel length $L = 5$ µm.

## Supplementary Note 4. *v* = 0 state in hBN/graphene/hBN

To demonstrate the *v* = 0 state observed in the transport measurement, we perform a detailed study of the temperature and magnetic field dependence of the $\rho_{xx}$ at $B = 0$. Two regions are identified in Supplementary Figure 6: $\rho_{xx}$ increases with decreasing temperature when the gate voltage is close to the DP, another region shows metallic behavior. From 250 K to 8.8 K, $\rho_{xx}$ at the DP approaches to the insulating state (Supplementary Figure 7a). With increasing $B$, a well-defined plateau of $\sigma_{xy}$ appears at the DP (Supplementary Figure 7b). At $B = 2.5$ T, $\rho_{xx}$ at the DP is close to 1 MΩ (Supplementary Figure 7d) and a double-peak structure in $\sigma_{xx}$ appears with $\sigma_{xy} = 0$ plateau (Supplementary Figure 7c). Supplementary Figure 7e, f show the evolution of $\sigma_{xx}$ and $\rho_{xx}$ with gate voltage and magnetic field.

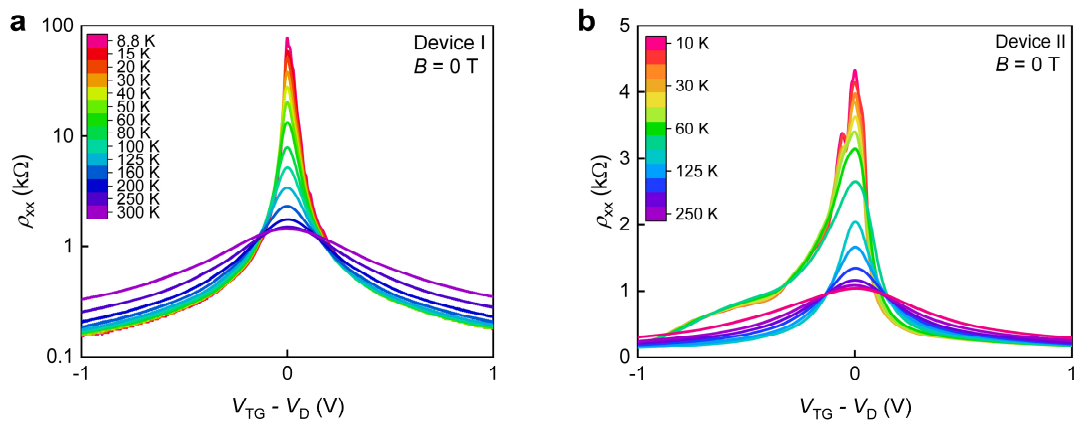

**Supplementary Fig. 6. Temperature dependence of local transport measurements.** Longitudinal resistivity ($\rho_{xx}$) vs gate voltage ($V_{TG} - V_D$) from 8.8 K to 300 K at zero magnetic field for devices I (**a**) and II (**b**).

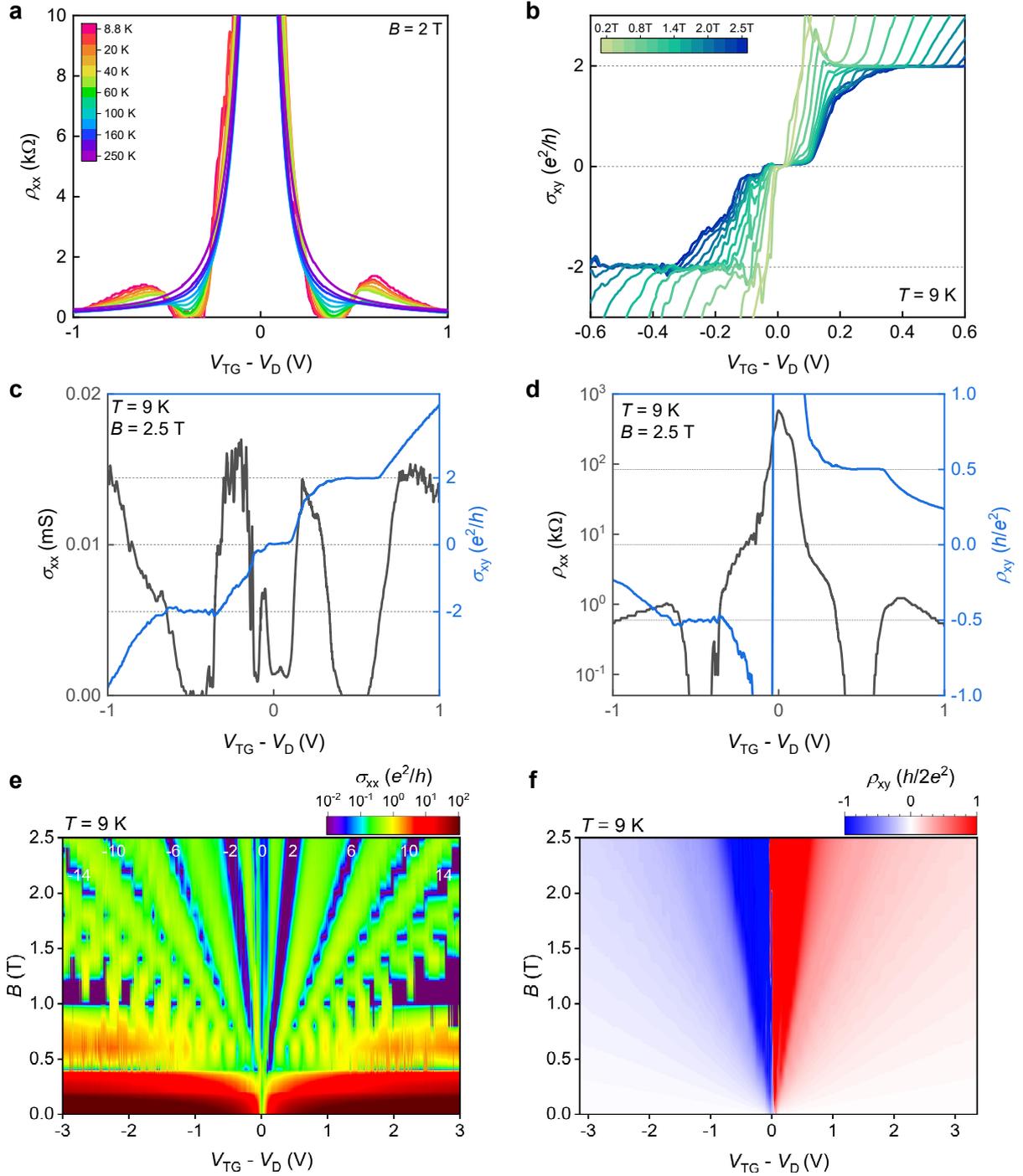

**Supplementary Fig. 7. ν = 0 state in device I. a** Temperature dependence of longitudinal resistivity ($\rho_{xx}$) vs gate voltage ($V_{TG} - V_D$) at 2 T. **b** Hall conductivity ($\sigma_{xy}$) vs gate voltage ($V_{TG} - V_D$) with magnetic field ($B$) from 0.2 T to 2.5 T at 9 K. **c** Longitudinal conductivity ($\sigma_{xx}$) and $\sigma_{xy}$, **d** $\rho_{xx}$ and Hall resistivity ($\rho_{xy}$) as a function of $V_{TG} - V_D$ at 2.5 T and 9 K. **e** $\sigma_{xx}$ and **f** $\rho_{xy}$ as functions of $V_{TG} - V_D$ and $B$ at 9 K. Numbers denote filling factors for the quantum Hall states extending from the Dirac point.

## Supplementary Note 5. Ohmic and thermal contributions to $R_{nl}$

Ohmic contribution is described by the van der Pauw formula[6],

$$R_{nl,\Omega} = \frac{W}{\pi L} R_{xx} \ln\left[\frac{\cosh(\pi L/W)+1}{\cosh(\pi L/W)-1}\right] \quad (S3)$$

where $L$ and $W$ are the channel length and width. In zero magnetic field for device I, $L/W = 2$ and $R_{xx}$ = 126 kΩ, and from Supplementary Equation (S3) we find $R_{nl,\Omega} \approx$ 150 Ω, which is two orders of magnitude smaller than the measured $R_{nl}$ in Supplementary Figure 8. In all transport measurements we use a low alternating-current excitation of 10 nA with a frequency 7 Hz. These low current amplitudes are chosen to minimize thermal contributions to the nonlocal transport due to Joule heating and Ettingshausen effects[7] whilst simultaneously maximizing the signal-to-noise ratio of the measured voltages. In zero magnetic field, only Joule heating effect contributes to the second harmonic of nonlocal signal $R_{nl,J}^{2f}$, which is less than 1% of $R_{nl}$ as shown in Supplementary Figure 8.

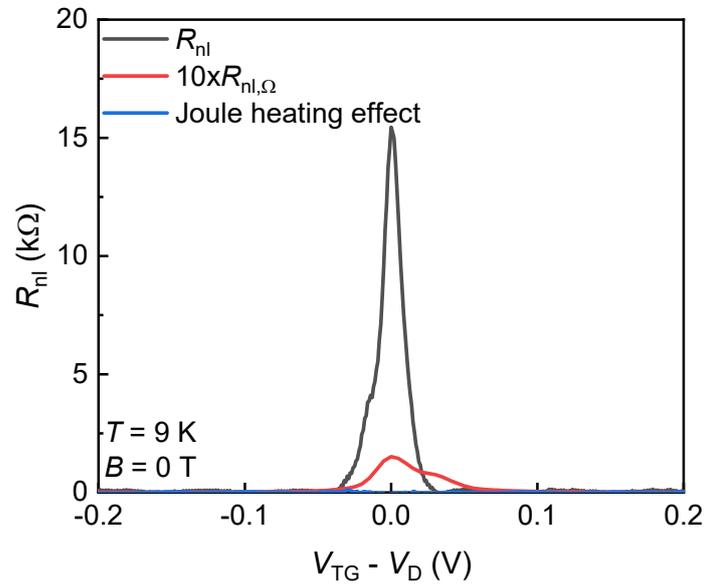

**Supplementary Fig. 8. Comparison of ohmic and Joule heating contributions to nonlocal resistance ($R_{nl}$).**

**Supplementary Note 6. Band gap calculation from nonlocal transport**

From the Arrhenius plot of the $R_{nl}$ (Supplementary Figure 9), the associated band gap is estimated as 760.4 ± 69.9 K assuming $1/R_{nl} \propto \exp(-E_a/2k_B T)$ (Ref. [8]). The obtained band gap is 2.3 times larger than calculated from the local transport (330.1 ± 18.3 K). However, this result is considered to be reasonable in the case that the $R_{nl} \propto \rho_{xx}^3$ in the high temperature regime ($T > 60$ K).

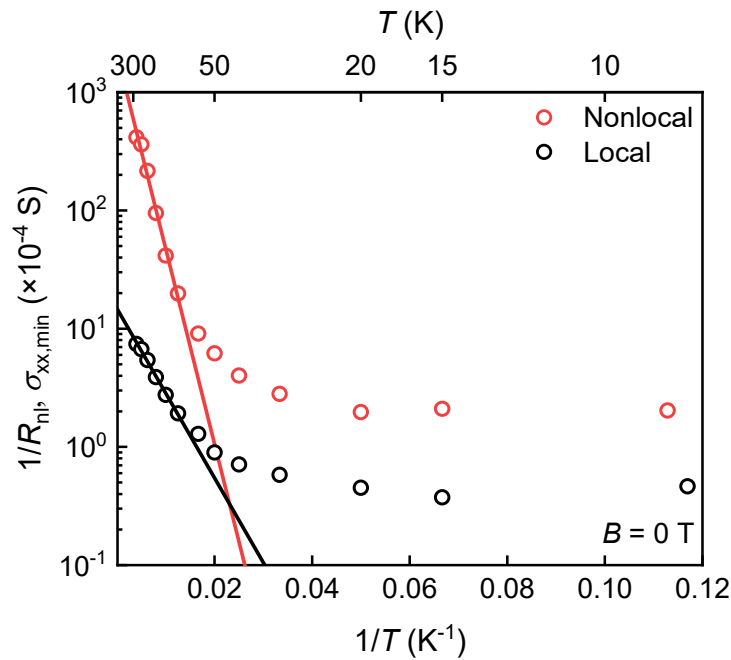

**Supplementary Fig. 9. Comparison of Arrhenius plots of nonlocal conductance ($1/R_{nl}$) and longitudinal conductivity ($\sigma_{xx}$) for device I.**

**Supplementary Note 7. Calculation of the transmission probability in the ten-terminal configuration**

According to the Landauer-Büttiker formalism, the current from terminal $p$ towards the device is given by

$$I_p = G_0 \sum_q Tr_{p,q}(V_p - V_q) \tag{S4}$$

where $G_0 = e^2/h$, and $Tr_{p,q} = Tr_{p \leftarrow q}$ is the transmission probability from terminal $q$ to $p$. We can express this in the matrix form as

$$\mathbf{I} = \mathbf{GV} \tag{S5}$$

where the conductance matrix $\mathbf{G}$ is given by

$$\mathbf{G} = G_0 \begin{pmatrix} \sum_q Tr_{1,q} & -Tr_{1,2} & -Tr_{1,3} & \cdots & -Tr_{1,10} \\ -Tr_{2,1} & \sum_q Tr_{2,q} & -Tr_{2,3} & \cdots & -Tr_{2,10} \\ -Tr_{3,1} & -Tr_{3,2} & \sum_q Tr_{3,q} & \cdots & -Tr_{3,10} \\ \vdots & \vdots & \vdots & \vdots & \cdots \\ -Tr_{10,1} & -Tr_{10,2} & -Tr_{10,3} & \cdots & \sum_q Tr_{10,q} \end{pmatrix} \tag{S6}$$

and **I** is the matrix where column describes a given current configuration (current injects from terminal to the device) and **V** is a matrix where each column describes the corresponding voltage configuration (voltages of different terminals).

These equations can be solved in three ways:

(i) If the voltage configuration and $Tr_{p,q}$ are known, we can directly determine the corresponding currents from Supplementary Equation (S5, S6).

(ii) If the current configuration and $Tr_{p,q}$ are known, we can determine the corresponding voltages in different terminals.

(iii) If we measure the ten independent current configuration with corresponding voltage, the conductance matrix can be determined from

$$\mathbf{G} = \mathbf{IV}^{-1} \tag{S7}$$

After determining **G**, the transmission matrix can be obtained from Supplementary Equation (S6). One possible way of obtaining ten independent current with corresponding voltages is to apply current between terminals $p-1$ and $p$, then cyclically shift the source and drain terminals so that one gets a 10×10 current matrix,

$$\mathbf{I} = I_0 \begin{pmatrix} -1 & 1 & 0 & \cdots & 0 \\ 0 & -1 & 1 & \cdots & 0 \\ \vdots & \ddots & -1 & \ddots & \vdots \\ \vdots & \ddots & \ddots & \ddots & 1 \\ 1 & 0 & 0 & \cdots & -1 \end{pmatrix} \tag{S8}$$

where $I_0$ is the magnitude of the current. For each current configuration, one measures the corresponding voltages in different terminals and gets voltage matrix **V**.

In a simple edge mode transport model proposed in Ref. [8] for quantum valley Hall state, a pair of ballistic edge modes connect terminal. Assuming that all terminals are ordered along the edge so that terminal $i$ is connected to both terminals $i-1$ and $i+1$, then $Tr_{ij}$ is given by

$$\mathbf{Tr} = \begin{pmatrix} 0 & 1 & 0 & \cdots & 1 \\ 1 & 0 & 1 & \cdots & 0 \\ 0 & 1 & \ddots & \ddots & \vdots \\ \vdots & \ddots & \ddots & \ddots & 1 \\ 1 & 0 & \cdots & 1 & 0 \end{pmatrix} \tag{S9}$$

For the current configuration given by Supplementary Equation (S8), the simple edge mode model predicts voltage matrix as

$$\mathbf{V} = \frac{I_0}{G_0} \begin{pmatrix} 0 & 0.9 & 0.8 & \cdots & 0.1 \\ 0.1 & 0 & 0.9 & \cdots & 0.2 \\ 0.2 & 0.1 & 0 & \ddots & \vdots \\ \vdots & \vdots & \ddots & \ddots & 0.9 \\ 0.9 & 0.8 & \cdots & 0.1 & 0 \end{pmatrix} \tag{S10}$$

Moreover, if the current configuration in Supplementary Equation (S8) and voltage configuration in Supplementary Equation (S10) are measured in the experiment, the conductance matrix can be determined from Supplementary Equation (S7) and then the transmission probabilities from Supplementary Equation (S6). We apply the above method for the measured $R_{nl}$ of ten-terminal configuration shown in Fig. 4e.

**Supplementary Table 1. Maximum transmission probabilities $Tr_{ij}$ in the range of gate voltage (-1.7 V ≤ $V_{TG}$ ≤ -1.5 V) for nonlocal resistance ($R_{nl}$) in the ten-terminal configuration.** $Tr_{ij}$ of terminals connected by edges are indicated with green color (shortest distance between terminals), blue color (intermediate distance between terminals) and red color (longest distance between terminals). Other pairs of terminals with $Tr_{ij}$ ≥ 0.5 are indicated with yellow color. The error bars for the transmission probabilities are ±0.2.

|     | T27 | T05 | T07 | T11 | T13 | T28 | T14 | T12 | T08 | T06 |
|-----|-----|-----|-----|-----|-----|-----|-----|-----|-----|-----|
| T27 |     | 1.0 | 0.7 | 0.1 | 0.1 | 0.2 | 0.2 | 0.2 | 0.6 | 1.3 |
| T05 | 1.1 |     | 0.4 | 0.1 | 0.0 | 0.1 | 0.1 | 0.2 | 0.4 | 1.1 |
| T07 | 0.7 | 0.5 |     | 0.1 | 0.1 | 0.3 | 0.2 | 0.4 | 0.8 | 0.6 |
| T11 | 0.1 | 0.1 | 0.2 |     | 0.2 | 0.5 | 0.5 | 0.6 | 0.1 | 0.1 |
| T13 | 0.1 | 0.1 | 0.1 | 0.2 |     | 0.7 | 0.7 | 0.5 | 0.1 | 0.1 |
| T28 | 0.2 | 0.2 | 0.5 | 0.6 | 1.0 |     | 1.0 | 0.7 | 0.2 | 0.1 |
| T14 | 0.1 | 0.1 | 0.3 | 0.3 | 0.6 | 1.3 |     | 1.8 | 0.2 | 0.1 |
| T12 | 0.2 | 0.1 | 0.2 | 0.4 | 0.4 | 0.8 | 2.0 |     | 0.3 | 0.1 |
| T08 | 0.6 | 0.3 | 0.8 | 0.2 | 0.1 | 0.2 | 0.2 | 0.3 |     | 0.6 |
| T06 | 1.3 | 1.0 | 0.6 | 0.1 | 0.1 | 0.2 | 0.1 | 0.1 | 0.4 |     |

The obtained transmission probabilities do not perfectly obey the time reversal symmetry ($Tr_{ij}$ = $Tr_{ji}$). It could be related to measurement errors, non-ideal contacts in-between the measurement terminals or weakly broken time reversal symmetry. The dominant values of $Tr_{ij}$ are marked with different colors (green, blue, red, and yellow) in Supplementary Table 1 and plotted in Fig. 4e. Because the distance between a certain numbers of terminals are longer than the intervalley diffusion length, we do not expect that the transmission probability given by Supplementary Equation (S9) would be perfectly reproduced. Nevertheless, if the nonlocal transport is dominated from edge modes, the values of $Tr_{ij}$ would obey the order of $Tr_{ij,\text{green}}$ > $Tr_{ij,\text{blue}}$ > $Tr_{ij,\text{red}}$. The transmission matrix does not strictly meet these relations and it seems that several terminals are connected in a more complicated way. However, the dominant contribution is still from spin-degenerate ballistic counter-propagating edge modes as discussed in the main text.

**Supplementary References**